# Understanding the electronic and phonon transport properties of thermoelectric material BiCuSeO: a first-principles study

D. D. Fan, H. J. Liu[*], L. Cheng, J. Zhang, P. H. Jiang, J. Wei, J. H. Liang, J. Shi

*Key Laboratory of Artificial Micro- and Nano-Structures of Ministry of Education and School of Physics and Technology, Wuhan University, Wuhan 430072, China*

Using first-principles pseudopotential method and Boltzmann transport theory, we give a comprehensive understanding of the electronic and phonon transport properties of thermoelectric material BiCuSeO. By choosing proper hybrid functional for the exchange-correlation energy, we find that the system is semiconducting with a direct band gap of ~0.8 eV, which is quite different from those obtained previously using standard functionals. Detailed analysis of a three-dimensional energy band structure indicates that there is a valley degeneracy of eight around the valence band maximum, which leads to a sharp density of states and is responsible for a large *p*-type Seebeck coefficient. Moreover, we find that the density of states effective masses are much larger and results in very low hole mobility of BiCuSeO. On the other hand, we find larger atomic displacement parameters for the Cu atoms, which indicates that the stronger anharmonicity of BiCuSeO may originate from the rattling behavior of Cu instead of previously believed Bi atoms.

## I. INTRODUCTION

Thermoelectric (TE) materials which can directly convert heat into electricity and vice versa have attracted much attention in the material science community because of increasing energy crisis and environmental pollution. The conversion efficiency of a TE material can be characterized by the dimensionless figure-of-merit $ZT = S^2\sigma T/(k_l + k_e)$, where $S$, $\sigma$, $T$, $k_e$, $k_l$ are the Seebeck coefficient, the electrical conductivity, the absolute temperature, the electronic thermal conductivity and the lattice thermal conductivity, respectively. A good TE material has a higher *ZT* value, which requires a larger power factor ($S^2\sigma$) and/or a lower thermal conductivity ($k_l + k_e$). However, it is usually difficult to do so since the transport coefficients $S$, $\sigma$ and $k_e$ are coupled with each other and related to the electronic band structure and carrier concentration. An effective way of searching good TE materials is to find materials with intrinsic low lattice thermal conductivity, such as CoSb$_3$ [1], AgSbTe$_2$ [2], and MgAgSb [3]. Recently, a new layered oxyselenide BiCuSeO with very small lattice thermal conductivity was proposed to be a potential TE material [4], and a large amount of efforts have been devoted to enhance its TE performance [5-9]. Over the past five years, the *ZT* value of BiCuSeO has been continuously increased from 0.76 to 1.5 by doping to enhance the power factor as well as nanostructuring to further reduce the lattice thermal conductivity. Many experimental works on the transport properties of BiCuSeO indicate that the compound exhibit (1) a relatively larger Seebeck coefficient at optimized concentration (230 $\mu$V/K @873K [9], 243 $\mu$V/K @923K [10]), (2) a lower mobility upon doping (generally lower than 5 cm$^2$/Vs [6, 7, 8, 11]), and (3) an intrinsic low lattice thermal conductivity (0.40 W/mK@923K [5], 0.22 W/mK@923K [8], 0.20 W/mK@873K [9]). In order to obtain a higher *ZT* value of BiCuSeO, it is crucial to have a better understanding of these transport properties. It was suggested by Zhao *et al.* [4] that the large Seebeck coefficient of BiCuSeO might be due to the two-dimensional confinement of the charge carriers in (Bi$_2$O$_2$)$^{2+}$ and (Cu$_2$Se$_2$)$^{2-}$ layers, which is very similar to other layered TE materials such as La$_2$CuO$_4$ [12] and Ca$_2$Co$_2$O$_5$ [13]. Moreover, the low lattice thermal conductivity may originate from the interface phonon scattering and low-phonon-conductive heavy elements [4]. Pei *et al.* [10] made a detailed analysis of the lattice thermal conductivity of BiCuSeO and found that it has a low Young's modulus (76.5 GPa) and sound velocity (2107 m/s), as well as a large Gruneisen parameter (~1.5) which indicates a strong anharmonicity. As for the origin of the low lattice thermal conductivity, it was speculated that the lone pair electrons of Bi$^{3+}$ possibly lead to a more asymmetric electron density and stronger lattice vibration energy [10]. Apart from the intrinsic phonon-phonon interactions, the grain boundary and defect were also believed to be important scattering mechanisms [6, 9, 10, 14]. In the theoretical aspect, the electronic band structure of BiCuSeO was studied by performing first-principles calculations [15], which suggests that a mixture of heavy and light bands near the valence band edge can lead to a relatively higher Seebeck coefficient and a reasonable electrical conductivity. A comparative study [16] of BiCuSeO and its isostructural analog LaCuSeO indicates that Bi atom with higher atomic mass can produce lower phonon frequencies, smaller sound velocities, and a higher average mode Gruneisen parameter. Ding *et al.* [17] investigated the structural scattering effect in BiCuSeO and revealed that the large Gruneisen parameter is caused by the strong anharmonic bonding of the heavy Bi and abnormal atomic motion originated from the layered structure. Using first-principles calculations, Shao *et al.* [18] showed that there is a strong bonding anharmonicity in BiCuSeO, leading to an ultralow lattice thermal conductivity. It should be noted that most of these works attribute the low thermal conductivity to the large atomic mass and the lone pair electrons of Bi atom. However, Vaqueiro *et al.* [19] argued that the major contribution to the low thermal conductivity may be the weak bonding of Cu atoms within the structure, rather than the previously believed $Bi^{3+}$ lone pair electrons. To summarize, the unique transport properties of BiCuSeO and the underlying physical mechanism are still under debate and a complete understanding is thus quite necessary.

---

[*] Author to whom correspondence should be addressed. Electronic mail: phlhj@whu.edu.cn

In this work, we show by first-principles calculations and Boltzmann transport theory that the band structure of BiCuSeO exhibits multiple valley degeneracy with a large density of states (DOS) effective mass, which leads to relatively higher Seebeck coefficient and lower mobility. Moreover, we discover several flat phonon modes dominated by the Cu and Se atoms, which can effectively block heat transfer in the $(Cu_2Se_2)^{2-}$ layer. Combined with large atomic displacement parameters (ADP) of Cu atom, we conclude that the intrinsic low lattice thermal conductivity in BiCuSeO is caused by Cu atoms, instead of previously believed Bi atoms.

## II. ELECTRONIC TRANSPORT

## A. Computational methods

The electronic properties of BiCuSeO are investigated within the framework of density functional theory (DFT) by using the projector-augmented wave (PAW) method, as implemented in the Vienna *ab-initio* simulation package (VASP) [20-22]. The exchange-correlation energy is in the form of Perdew-Burke-Ernzerhof (PBE) with generalized gradient approximation (GGA) [23]. To accurately predict the band gap, the hybrid density functional using the form of Heyd-Scuseria-Ernzerhof (HSE, version 06) is also employed [24-26] and compared with standard DFT calculations. The plane-wave cutoff is set to 400 eV and the energy convergence threshold is $10^{-6}$ eV. The spin-orbital coupling (SOC) is explicitly included in the calculations. The Seebeck coefficient is evaluated by using the semiclassical Boltzmann transport theory with the relaxation time approximation, which is implemented in the so-called BoltzTraP code [27].

## B. Band structure

The BiCuSeO compound crystallizes in a tetragonal structure with space group *P4/nmm* [28]. Figure 1(a) displays the top-view of the crystal structure, where we see obvious fourfold rotational symmetry along the *c*-axis direction. The structure can be viewed as alternatively stacking the $(Bi_2O_2)^{2+}$ and $(Cu_2Se_2)^{2-}$ layers, as shown in Fig. 1(b). The optimized lattice parameters of BiCuSeO are $a = b = 3.956$ Å and $c = 9.108$ Å, which are slightly larger than the experimental results [28].

Figure 2 plots the band structures of BiCuSeO along several high symmetry lines in the irreducible Brillion zone, where the results with PBE and HSE functionals are both shown. For the PBE calculations, we see from Fig. 2(a) that the SOC tends to upshift the valence bands and downshift the conduction bands. This is especially pronounced for the energy bands around the *Z* point which tends to close the band gap. Moreover, the effect of SOC also changes the location of valence band maximum (VBM) from that between the *M* and *Γ* points to the *Z* point, which is different from previous theoretical predictions [15, 29, 30]. Similar picture can be found in Fig. 2(b) by using the HSE functional, except that there is a direct band gap of 0.81 eV which is very close to experimental results [29-31]. In addition to the VBM, there are two other valence band extremums (VBEs) with almost identical energies, one appears between the *M* and *Γ* points, and the other is located between the *Z* and *R* points. As these valence valleys have comparable energies (Table I), it is very possible that a slight difference in the lattice parameters or adopting different exchange-correlation functionals may change the position of the VBM, as discussed above. It should be emphasized that the measured optical absorption spectra [29] of BiCuSeO shows sharp band-edge structures, which gives obvious evidence of a direct band gap, as predicted by our hybrid functional calculations with SOC explicitly considered.

### C. DOS effective mass and mobility

To understand why BiCuSeO has a relatively lower mobility as compared with other well-known thermoelectric materials [32-36], we show in Figure 3(a) and 3(b) the energy dispersion relations of the top valence band at $k_Z$=0.0 and 0.5 planes, respectively. Along the *ΓM* and *ZR* directions, the effective mass of the VBE is respectively calculated to be 0.23 $m_e$ and 4.53 $m_e$, which shows the features of light and heavy bands and is consistent with previous analysis [15]. However, the case is just reversed for the directions perpendicular to the *ΓM* and *ZR* directions, where the VBE has an effective mass of 6.79 $m_e$ and 0.45 $m_e$, respectively. Such anisotropic effective mass indicates that we should consider the DOS effective masses, which are calculated to be $m_1$=0.61 $m_e$ for the VBM, $m_2$=1.05 $m_e$ for the VBE shown in Fig. 2(a), and $m_3$=1.30 $m_e$ for the VBE shown in Fig. 2(b). The total DOS effective mass is calculated to be 4.9 $m_e$ for the top valence band by using the formula $m^*_{total} = (N_1 m_1^{3/2} + N_2 m_2^{3/2} + N_3 m_3^{3/2})^{2/3}$, where $N_1$=1, $N_2$=4, and $N_3$=4 are the valley degeneracy for the VBM and two VBEs, respectively. The averaged DOS effective mass of one valley, given by $m^*_{dos} = N^{-2/3} m^*_{total}$, is obtained by considering a total valley degeneracy of *N*=8 (we will come back to this point later). The calculated $m^*_{dos}$ is 1.22 $m_e$, which is quite large and implies rather small carrier mobility as governed by $\mu \propto m_{dos}^{*-5/2}$ [37]. For comparison, we show in Fig. 3(c) the extracted DOS effective mass from some experimental results [6, 38-40] by employing the widely used parabolic band model [41]. When the carrier concentration is below a certain value, we see the DOS effective mass is small, which corresponds to the case that the Fermi level only crosses the VBM (left inset). When the Fermi level enters into the two VBEs at higher carrier concentration, there is an obvious increase of the DOS effective mass. By making an average in this concentration region, the DOS effective mass is estimated to be 1.23 $m_e$, which is almost the same as our theoretically predicted value.

### D. Multiple valley degeneracy

As can be clearly seen from Fig. 3(a) and 3(b), that there are four equivalent valleys for each VBE of BiCuSeO, which can be attributed to the fourfold rotational symmetry indicated in Fig. 1(a). Since the energy difference between the two VBEs is quite small (See Table I), there actually exists a valley degeneracy of eight around the VBM. To have a better understanding, we plot in Fig. 3(d) the constant energy surface near VBM within the first Brillouin zone. Apart from the four equivalent valleys in the middle of the zone, the remaining four equivalent valleys are separated into two parts at the top and bottom of the zone. It is interesting to note that the band structures of some state-of-the-art TE materials such as $Bi_2Te_3$ [37] and PbTe [42] also exhibit feature of multiple valley degeneracy, which can be used as an important paradigm in searching high performance TE materials [43].

### E. DOS and Seebeck coefficient

Figure 4(a) plots the partial DOS of BiCuSeO, where we see that the DOS around the VBM are mainly determined by the Cu and Se atoms. Moreover, there is a sharp increase of the DOS with the decrease of energy, which suggests a large *p*-type Seebeck coefficient of BiCuSeO according to the modified Mott formula [44]. This is also consistent with the large DOS effective mass and the multiple valley degeneracy discussed above. Indeed, we see from Fig. 4(b) that the room temperature Seebeck coefficient [6, 38-40, 45-47] can be as large as ~400 μV/K. Note that our calculated Seebeck coefficients almost coincide with the experimental data at high carrier concentration, which further confirms the reliability of our theoretical approach. The discrepancy at low carrier concentration region may be caused by the bipolar effect, where the contribution from minority carriers makes some cancel of the Seebeck coefficient. On the other hand, the DOS contributed from Bi element is very small, which may be utilized to enhance the thermoelectric performance of BiCuSeO by doping at the Bi site with less change in the DOS but effectively reducing the lattice thermal conductivity [48].

## III. PHONON TRANSPORT

### A. Computational method

The phonon dispersion relations of BiCuSeO are calculated using the supercell approach with finite displacement as implemented in the Phonopy package [49]. To obtain the second order force constants (2FC), a $3\times 3\times 2$ Monkhorst-Pack $k$-point mesh is adopted and a $3\times 3\times 2$ supercell with 312 structures is employed. The phonon transport properties can be obtained by solving the phonon Boltzmann transport equation as implemented in the so-called ShengBTE code [50]. The third order force constants (3FC) are calculated by a finite difference approach using a $3\times 3\times 2$ supercell with totally 752 structures. The eighth nearest neighbors are included for the third-order interactions as suggested by Shao *et al.* [18]. A $15\times 15\times 7$ $q$-mesh is adopted to get a converged lattice thermal conductivity.

### B. Phonon dispersion relations

To understand the intrinsic low lattice thermal conductivity of BiCuSeO, we begin with the phonon dispersion relations as shown in Figure 5(a), where the projected phonon DOS is also given. It is found that there is a frequency gap around 200 cm$^{-1}$, above which the high-frequency optic phonon modes are almost entirely contributed by O atoms. Moreover, the large group velocity suggests a relatively bigger contribution of optic phonon to the lattice thermal conductivity. In the intermediate-frequency region, we observe two branches of phonon bands with less dispersions (shadow area). Such kind of flat modes are dominated by Se (upper branch) and Cu atoms (lower one), which can lead to much stronger anharmonic scattering between phonons [51]. In addition, we find in the low-frequency region that the acoustic phonon branches are mainly contributed by Bi atoms. Due to relatively larger group velocity and higher occupation possibility, these low-frequency phonon modes could have a major contribution to the heat transport. It should be noted that there are several low-frequency optic branches (near 60 cm$^{-1}$) mixed with the acoustic phonon modes, which is usually found in many low thermal conductivity materials containing heavy atoms [52]. All of these observations will be helpful to understand the phonon transport properties of BiCuSeO compound, as discussed in the following.

## C. Lattice thermal conductivity

The calculated lattice thermal conductivity of BiCuSeO is displayed in Figure 5(b) as a function of temperature. We see that the in-plane thermal conductivity ($k_a$) is much larger than the out-of-plane one ($k_c$), which is due to the weak bonding between the $(Bi_2O_2)^{2+}$ and $(Cu_2Se_2)^{2-}$ layers. Compared with the experimental results [4, 6, 10, 45, 53], the averaged lattice thermal conductivity $k_{ave} = (k_a + k_b + k_c)/3$ is somehow larger, which may be due to the fact that our calculations are done for a perfect single crystal while the samples used in the experiments are usually polycrystalline. To understand the effect of the flat modes shown in Fig. 5(a), we plot in Fig. 5(c) the phonon relaxation time $\tau_{ph}$ as a function of frequency. It is found that the Cu and Se dominated flat modes indeed have relatively lower relaxation time, which is caused by the enhanced three-phonon scattering. Besides, we find that the relaxation time of acoustic phonon is very large while that of the high-frequency optic phonon is much lower, which is consistent with the simple relation $\tau_{ph} \propto \omega^{-2}$ ($\omega$ is the phonon frequency) in the phonon Umklapp process [54, 55]. In Fig. 5(d), we show the normalized accumulative lattice thermal conductivity at room temperature with respect to cutoff phonon frequency. It is found that the lattice thermal conductivity increases quickly with $\omega$ in the low-frequency region. By setting a cutoff of 50 cm$^{-1}$, the normalized accumulated thermal conductivity is found to be as high as ~45%, which is due to the large group velocity, occupation possibility and relaxation time as discussed above. In the intermediate-frequency region from 50 to 150 cm$^{-1}$, we see that two flat modes contributes less to the thermal conductivity though they have a large phonon DOS. Such flat modes have a relatively small phonon relaxation time and group velocity, which are usually believed to be heat insulating. As discussed above, the high-frequency optic phonons are dominated by the O atoms, and contribute ~25% of the lattice thermal conductivity due to large group velocity.

### D. Origin of the anharmonicity

To figure out the origin of the intrinsic low lattice thermal conductivity of BiCuSeO compound, we first calculate the Gruneisen parameter and find a very large value of 2.5, which indicates a high degree of anharmonicity and a lower thermal conductivity. Note that the Gruneisen parameter was estimated to be 1.5 in Reference [10], which may be caused by the neglect of optic phonons. It was previously suggested that the Bi atom contributes to the large anharmonicity because of its large atomic mass and unique lone pair electrons [10, 16, 17]. To see if this is the real case, we calculate the ADP of BiCuSeO. Figure 6(a) plots the ADP of each atom as a function of temperature, which in principle shows similar behavior as those measured by neutron diffraction [19]. Moreover, we find that Cu atom has the largest ADP value and increases much faster than those of other atoms when the temperature is increased. It is thus reasonable to expect that the strong anharmonicity of BiCuSeO may originate from the Cu atoms. To further confirm this point, we conduct a first-principles molecular dynamic (MD) simulation to show the trajectory of each atom. We select a microcanonical ensemble and the temperature is set to 300 K. The MD runs for 30,000 steps with a time step of 0.5 fs. As shown in Fig. 6(b), there are only small vibrations of the Bi and O atoms around their equilibrium positions. In contrast, the rattling motion of Cu and Se atoms are much larger (especially for the Cu atoms), which indicates very strong anharmonicity. On the other hand, we have shown that Cu and Se atoms dominate two frequency regions with rather flat phonon modes (Fig. 5(a)). As a consequence, the heat transport may "break down" when passing through these atoms. Combing all these results, we believe that the intrinsic low lattice thermal conductivity is most likely to be caused by Cu atoms, rather than previously believed Bi atoms.

## IV. CONCLUSIONS

We present a comprehensive theoretical study on the electronic and phonon transport properties of TE material BiCuSeO within the framework of DFT. We obtain a direct band gap of ~0.8 eV by choosing the hybrid functional for the exchange-correlation energy and considering SOC. Detailed analysis of the band structure in the whole Brillouin zone reveals the features of large DOS effective mass and multiple valley degeneracy, which helps to understand the low mobility and high Seebeck coefficient of BiCuSeO. On the other hand, we demonstrate that the Cu and Se induced flat phonon modes can effectively block heat transfer. In particular, the Cu atom exhibits the largest ADP, giving direct evidence that the strong anharmonicity of BiCuSeO should originate from the Cu atom rather than previously believed Bi atom. It is interesting to note that the charge transport of BiCuSeO is mainly governed by the $(Cu_2Se_2)^{2-}$ layer, while the $(Bi_2O_2)^{2+}$ layer plays a major role in the heat transport. Such unique characteristic could be used to independently manipulate the electron and phonon transport so that the TE performance of BiCuSeO can be further enhanced.

## ACKNOWLEDGMENTS

We thank financial support from the National Natural Science Foundation (grant No. 11574236 and 51172167) and the "973 Program" of China (Grant No. 2013CB632502)

**Table I** The energy eigenvalues (in unit of eV) of VBM and two VBEs of BiCuSeO.

| VBM | VBE ($\Gamma M$ direction) | VBE ($ZR$ direction) |
|---|---|---|
| −0.438 | −0.482 | −0.489 |

**Table II** The effective mass (in unit of $m_e$) of three valence valleys along different directions for BiCuSeO. The corresponding DOS effective mass is also listed.

| | | | | |
|---|---|---|---|---|
| VBM | $m_{ZR}$ | $m_{\perp ZR}$ | $m_{k_z}$ | $m^*_{dos}$ |
| | 0.47 | 0.47 | 1.06 | 0.61 |
| VBE ($\Gamma M$ direction) | $m_{\Gamma M}$ | $m_{\perp \Gamma M}$ | $m_{k_z}$ | $m^*_{dos}$ |
| | 0.23 | 6.79 | 0.74 | 1.05 |
| VBE ($ZR$ direction) | $m_{ZR}$ | $m_{\perp ZR}$ | $m_{k_z}$ | $m^*_{dos}$ |
| | 4.53 | 0.45 | 1.09 | 1.30 |

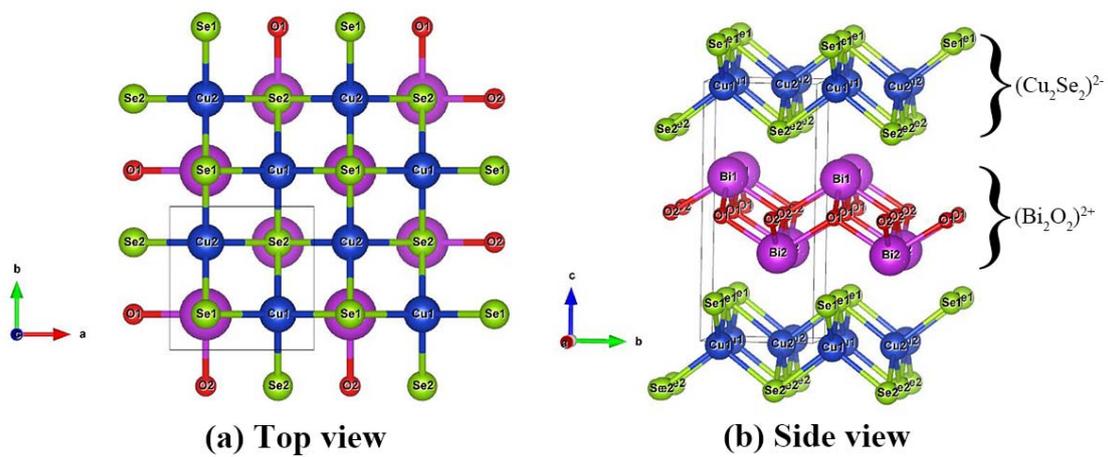

**Figure 1** Ball and stick models of BiCuSeO with (a) top-view and (b) side-view.

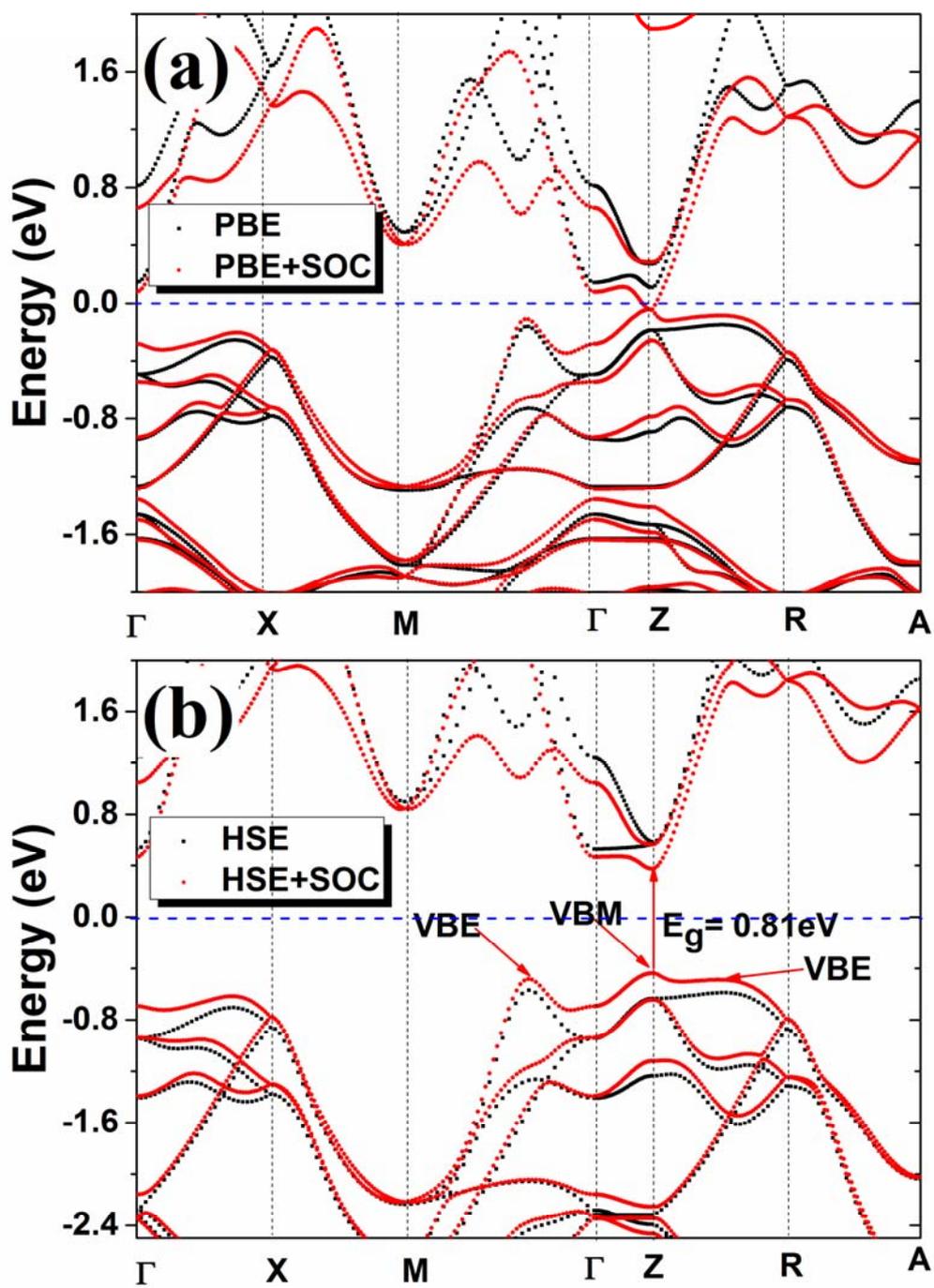

**Figure 2** Band structures of BiCuSeO calculated with (a) PBE and (b) HSE functionals. The red and black lines correspond to the calculations with and without SOC, respectively. The Fermi level is set at 0 eV.

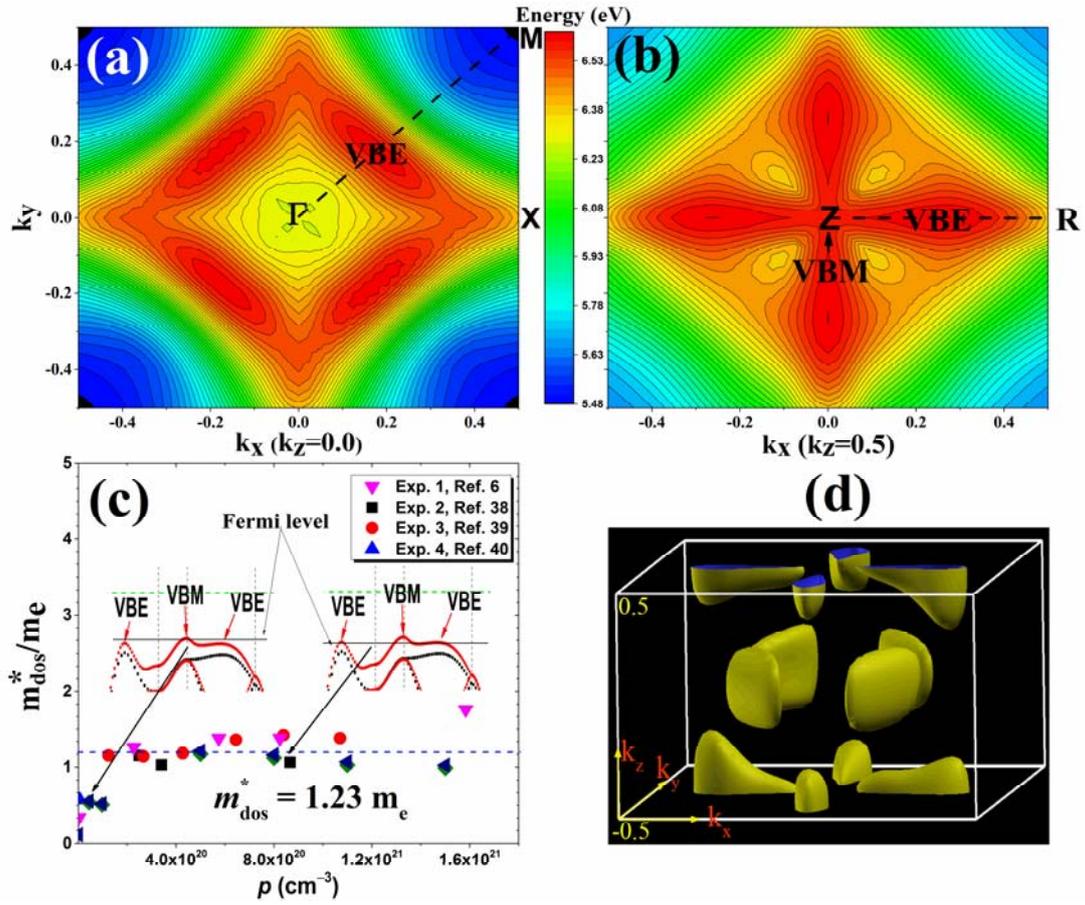

**Figure 3** The energy dispersion relations of the top valence band for BiCuSeO (a) at $k_Z$=0.0 plane and (b) at $k_Z$=0.5 plane, where the energy eigenvalues are denoted by different colors. (c) is the DOS effective mass of BiCuSeO derived from experimental data, plotted as a function of carrier concentration. The insets illustrate the position of Fermi level at different carrier concentration. (d) shows eight valence valleys in the first Brillouin zone.

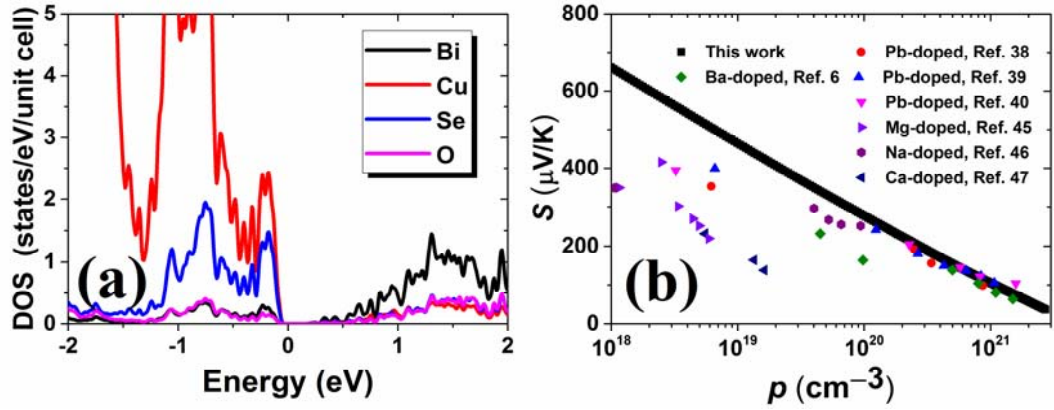

**Figure 4** (a) The DOS of BiCuSeO projected on each atom. The Fermi level is at 0 eV. (b) The room temperature Seebeck coefficient as a function of carrier concentration. The experimental data are marked for comparison.

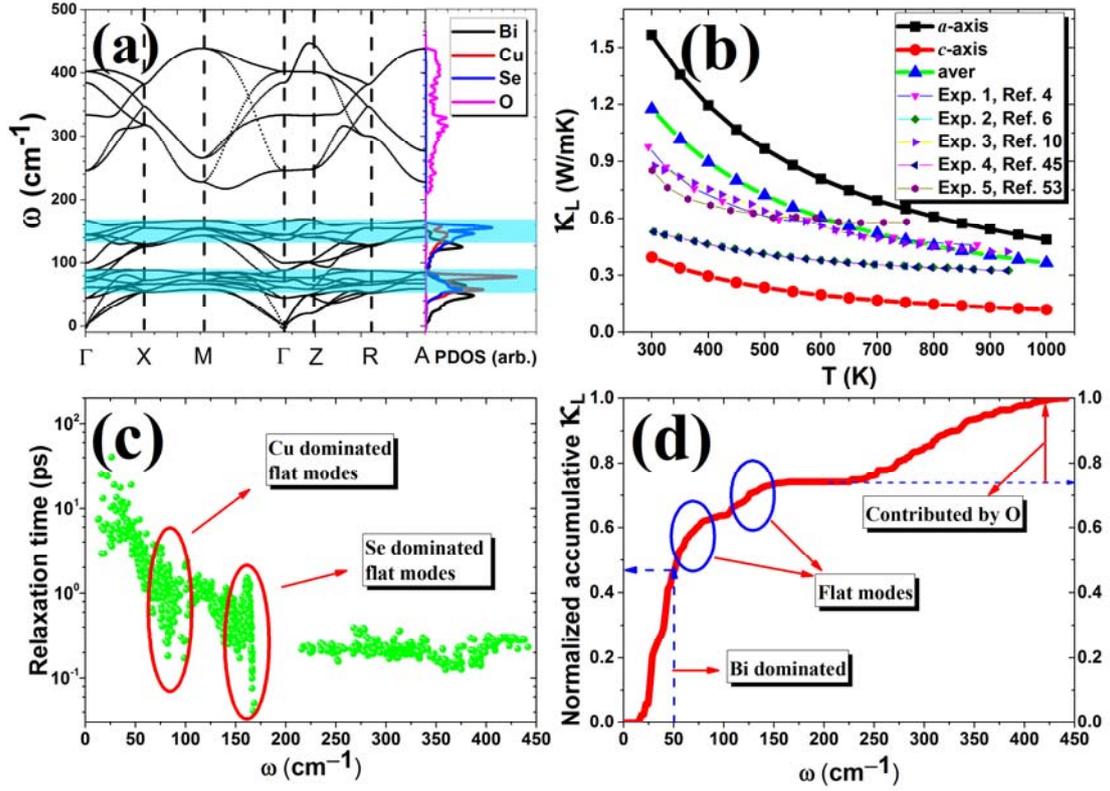

**Figure 5** (a) The phonon dispersion relations and projected DOS of BiCuSeO. (b) The calculated lattice thermal conductivity of BiCuSeO, plotted as a function of temperature and compared with available experimental data. (c) The phonon relaxation time of BiCuSeO as a function of frequency. (d) Normalized accumulative lattice thermal conductivity (at 300 K) with respect to the cutoff frequency.

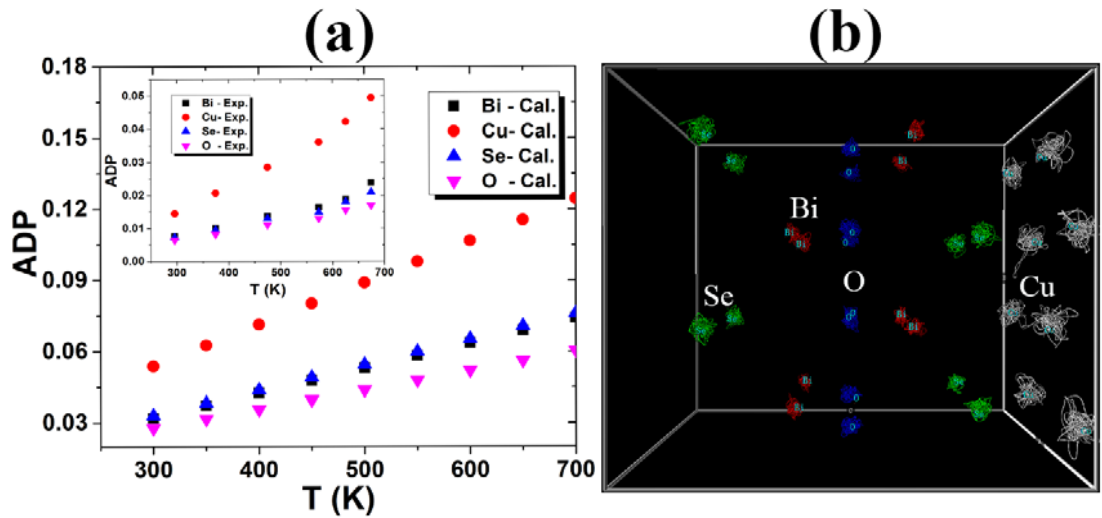

**Figure 6** (a) Calculated ADP of BiCuSeO as a function of temperature. The inset displays the experimental data measured by neutron diffraction. (b) First-principles MD simulations of BiCuSeO showing the trajectory of each atom.

**References**

[1] G. S. Nolas, D. T. Morelli and T. M. Tritt, Annu. Rev. Mater. Sci. **29**, 89 (1999).

[2] D. T. Morelli, V. Jovovic and J. P. Heremans, Phys. Rev. Lett. **101**, 035901 (2008).

[3] H. Z. Zhao, J. H. Sui, Z. J. Tang, Y. C. Lan, Q. Jie, D. Kraemer, K. McEnaney, A. Guloy, G. Chen, and Z. F. Ren, Nano. Energy **7**, 97 (2014).

[4] L. D. Zhao, D. Berardan, Y. L. Pei, C. Byl, L. Pinsard-Gaudart, and N. Dragoe, Appl. Phys. Lett. **97**, 092118 (2010).

[5] Y. Liu, L. D. Zhao, Y. C. Liu, J. L. Lan, W. Xu, F. Li, B. P. Zhang, David Berardan, N. Dragoe, Y. H. Lin, C. W. Nan, J. F. Li, and H. Zhu, J. Am. Chem. Soc. **133**, 20112 (2011).

[6] J. Li, J. H. Sui, Y. L. Pei, C. Barreteau, D. Berardan, N. Dragoe, W. Cai, J. Q. He, and L. D. Zhao, Energy Environ. Sci. **5**, 8543 (2012).

[7] J. H. Sui, J. Li, J. Q. He, Y. L. Pei, D. Berardan, H. J. Wu, N. Dragoe, W. Cai and L. D. Zhao, Energy Environ. Sci. **6**, 2916 (2013).

[8] Y. L. Pei, H. J. Wu, D. Wu, F. S. Zhang and J. Q. He, J. Am. Chem. Soc. **136**, 13902 (2014).

[9] Y. Liu, L. D. Zhao, Y. C. Zhu, Y. C. Liu, F. Li, M. J. Yu, D. B. Liu, W. Xu, Y. H. Lin, and C. W. Nan, Adv. Energy Mater. **6**, 1502423 (2016).

[10] Y. L. Pei, J. Q. He, J. F. Li, F. Li, Q. J. Liu, W. Pan, C. Barreteau, D. Berardan, N. Dragoe, and L. D. Zhao, NPG Asia Mater. **5**, 47 (2013).

[11] J. L. Lan, B. Zhan, Y. C. Liu, B. Zheng, Y. Liu, Y. H. Lin, and C. W. Nan, Appl. Phys. Lett. **102**, 123905 (2013).

[12] Y. Liu, Y. H. Lin, B. P. Zhang, H. M. Zhu, C. W. Nan, J. L. Lan and J. F. Li, J. Am. Ceram. Soc. **92**, 934 (2009).

[13] J. L. Lan, Y. H. Lin, G. J. Li, S. L. Xu, Y. Liu, C. W. Nan and S. J. Zhao, Appl. Phys. Lett. **96**, 192104 (2010).

[14] F. Li, T. R. Wei, F. Y. Kang and J. F. Li, J. Mater. Chem. A. **1**, 11942 (2013).

[15] D. F. Zou, S. H. Xie, Y. Y. Liu, J. G. Lin and J. Y. Li, J. Mater. Chem. A, **1**, 8888 (2013).

[16] S. K. Saha, Phys. Rev. B **92**, 041202 (2015).

[17] J. X. Ding, B. Xu, Y. H. Lin, C. W. Nan and W. Liu, New J. Phys. **17**, 083012 (2015).

[18] H. Z. Shao, X. J. Tan, G. Q. Liu, J. Jiang and H. C. Jiang, Sci. Rep. **6**, 21035 (2016).

[19] P. Vaqueiro, R. A. R. Al Orabi, S. D. N. Luu, G. Guelou, A. V. Powell, R. I. Smith, J. P. Song, D. Wee, and M. Fornari, Phys. Chem. Chem. Phys. **17**, 31735 (2015).

[20] G. Kresse and J. Hafner, Phys. Rev. B **47**, 558 (1993).

[21] G. Kresse and J. Hafner, Phys. Rev. B **49**, 14251 (1994).

[22] G. Kresse and J. Furthmüller, Comput. Mater. Sci. **6**, 15 (1996).

[23] J. P. Perdew, K. Burke and M. Ernzerhof, Phys. Rev. Lett. **77**, 3865 (1996).

[24] J. Heyd, G. E. Scuseria and M. Ernzerhof, J. Chem. Phys. **118**, 8207 (2003).

[25] J. Heyd and G. E. Scuseria, J. Chem. Phys. **121**, 1187 (2004).

[26] J. Heyd, G. E. Scuseria, and M. Ernzerhof, J. Chem. Phys. **124**, 219906 (2006).

[27] G. K. H. Madsen and D. J. Singh, Comput. Phys. Commun. **175**, 67 (2006).

[28] L. D. Zhao, J. Q. He, D. Berardan, Y. H. Lin, J. F. Li, C. W. Nan and N. Dragoe, Energy Environ. Sci. **7**, 2900 (2014).

[29] H. Hiramatsu, H. Yanagi, T. Kamiya, K. Ueda, M. Hirano and H. Hosono, Chem. Mater. **20**, 326 (2008).

[30] E. S. Stampler, W. C. Sheets, M. I. Bertoni, W. Prellier, T. O. Mason and K. R. Poeppelmeier, Inorg. Chem. **47**, 10009 (2008).

[31] A. P. Richard, J. A. Russell, A. Zakutayev, L. N. Zakharov, D. A. Keszler, and J. Tate, J. Solid State Chem. **187**, 15 (2012)

[32] Y. Z. Pei, Zachary M. Gibbs, A. Gloskovskii, B. Balke, W. G. Zeier, and G. J. Snyder, Adv. Energy Mater. **4**, 1400486 (2014).

[33] C. L. Chen, H. Wang, Y. Y. Chen, T. Day and G. J. Snyder, J. Mater. Chem. A **2**, 11171 (2014).

[34] J. H. Sui, J. Shuai, Y. C. Lan, Y. Liu, R. He, D. Z. Wang, Q. Jie and Z. F. Ren, Acta Materialia **87**, 266 (2015).

[35] R. Akram, Y. G. Yan, D. W. Yang, X. Y. She, G. Zheng, X. L. Su, and X. F. Tang, Intermetallics **74**, 1 (2016).

[36] X. H. Liu, T. J. Zhu, H. Wang, L. P. Hu, H. H. Xie, G. Y. Jiang, G. J. Snyder and X. B. Zhao, Adv. Energy Mater. **3**, 1238 (2013).

[37] H. J. Goldsmid, *Thermoelectric Refrigeration* (Plenum, 1964).

[38] J. L. Lan, Y. C. Liu, B. Zhan, Y. H. Lin, B. P. Zhang, X. Yuan, W. Q. Zhang, W. Xu, and C. W. Nan, Adv. Mat. **25**, 5086 (2013).

[39] L. Pan, D. Berardan, L. D. Zhao, C. Barreteau and N. Dragoe, App. Phys. Lett. **102**, 023902 (2013).

[40] G. K. Ren, J. L. Lan, S. Butt, K. J. Ventura, Y. H. Lin and C. W. Nan, RSC Adv. **5**, 69878 (2015).

[41] D. M. Rowe, C. M. Bhandari, *Modern thermoelectric*, Reston Publishing Company, Inc.: Reston Virginia, pp. 26 (1983).

[42] Y. Z. Pei, X. Y. Shi, A. LaLonde, H. Wang, L. D. Chen and G. J. Snyder, Nature **473**, 66 (2011).

[43] F. J. DiSalvo, Science **285**, 703 (1999).

[44] J. P. Heremans, V. Jovovic, E. S. Toberer, A. Saramat, K. Kurosaki, A. Charoenphakdee, S. Yamanaka, and G. J. Snyder, Science **321**, 554 (2008).

[45] J. Li, J. H. Sui, C. Barreteau, D. Berardan, N. Dragoe, W. Cai, Y. L. Pei, and L. D. Zhao, J. Alloy Compd. **551**, 649 (2013).

[46] J. Li, J. H. Sui, Y. L. Pei, X. F. Meng, D. Berardan, N. Dragoe, W. Cai and L. D. Zhao, J. Mater. Chem. A **2**, 4903 (2014).

[47] C. L. Hsiao and X. D. Qi, Acta Materialia **102**, 88 (2016).

[48] P. F. P. Poudeu, J. D'Angelo, H. J. Kong, A. Downey, J. L. Short, R. Pcionek, T. P. Hogan, C. Uher, and M. G. Kanatzidis, J. Am. Chem. Soc. **128**, 14347 (2006).

[49] A. Togo and I. Tanaka, Scr. Mater. **108**, 1 (2015)

[50] W. Li, J. Carrete, N. A. Katcho, and N. Mingo, Comput. Phys. Commun. **185**, 1747 (2014).

[51] W. Li and N. Mingo, Phys. Rev. B, **91**, 144304 (2015).

[52] D. Wee, B. Kozinsky, N. Marzari and M. Fornari, Phys. Rev. B **81**, 045204 (2010).

[53] Z. Li, C. Xiao, S. J. Fan, Y. Deng, W. S. Zhang, B. J. Ye and Y. Xie, J. Am. Chem. Soc. **137**, 6587 (2015).

[54] P. G. Klemens, J. Wide Bandgap Mater. **7**, 332 (2000).

[55] P. G. Klemens, Int. J. Thermophys. **22**, 265 (2001).